\documentstyle[aps,prd,multicol,epsfig]{revtex}

\begin{document}

\title{Numerical and Analytical Predictions for the Large-Scale
Sunyaev-Zel'dovich Effect}

\author{Alexandre Refregier}
\address{Institute of Astronomy, Madingley Road, University of
Cambridge, Cambridge CB3 OHA, England; ar@ast.cam.ac.uk}

\author{Romain Teyssier}
\address{Service d'Astrophysique, Bat. 709, CEA Saclay, F-91191
Gif sur Yvette, France; Romain.Teyssier@cea.fr}

\date{Submitted to PRD - December 4, 2000}

\maketitle

\begin{abstract}
The hot gas embedded in the large-scale structures in the Universe
produces secondary fluctuations in the Cosmic Microwave Background
(CMB). Because it is proportional to the gas pressure integrated along
the line of sight, this effect, the thermal Sunyaev-Zel'dovich (SZ)
effect, provides a direct measure of large-scale structure and of
cosmological parameters.  We study the statistical properties of this
effect using both hydrodynamical simulations and analytical
predictions from an extended halo model.  The Adaptive Mesh Refinement
scheme, used in the newly developed code RAMSES, provides a dynamic
range of 4 order of magnitudes, and thus allows us to significantly
improve upon earlier calculations.  After accounting for the finite
mass resolution and box size of the simulation, we find that the halo
model agrees well with the simulations. We discuss and quantify the
uncertainty in both methods, and thus derive an accurate prediction
for the SZ power spectrum in the $10^{2} \lesssim l \lesssim 10^5$
range of multipole. We show how this combined analytical and numerical
approach is essential for accuracy, and useful for the understanding
of the physical processes and scales which contribute to large-scale
SZ anisotropies.
\end{abstract}


\section{Introduction}

The hot gas in the IGM induces distortions in the spectrum of the
Cosmic Microwave Background (CMB) through the Sunyaev-Zel'dovich (SZ)
effect \cite{sun72,sun80}. This effect is proportional to the
integrated electron pressure along the line-of-sight and can be
directly measured by mapping the anisotropies of the CMB (see
\onlinecite{rep95,bir99} for reviews). It acts as a foreground for the
measurement of primary anisotropies of the CMB (see eg.
\onlinecite{ref99a} for a review), and is a direct probe of large-sale
structure and of the distribution of baryons in the universe. This
technique is now well established for the study of individual clusters
of galaxies (eg.  \onlinecite{rep95,bir99,car99,gra99}). A more
complete and unbiased statistical description of the SZ effect
requires wide field surveys, and will soon be achieved by upcoming and
future CMB experiments (see \cite{agh97,ref00a,coo+00} and reference
therein). In this perspective, accurate theoretical predictions of the
statistics of the large-scale SZ effect are necessary.

The statistics of SZ anisotropies have been studied using various
methods. Scaramella et al. \cite{sca93}, and more recently da Silva
et al. \cite{das99,das00} and Seljak, Burwell \& Pen \cite{sel99},
have used numerical simulations to construct SZ maps and study
their statistical properties.  Persi et al. \cite{per95} and
Refregier et al. \cite{ref00} used instead a semi-analytical
method, consisting of computing the SZ angular power spectrum by
projecting the 3-dimensional power spectrum of the gas pressure on
the sky. Aghanim et al. \cite{agh97}, Atrio-Barandela \& M\"ucket
\cite{atr99} and Komatsu \& Kitayama \cite{kom99} used the Press \&
Schechter formalism \cite{pre74} to predict the SZ power spectrum
in various CDM models. In the same spirit, Cooray \cite{coo00}
recently used a self-consistent extended halo model
\cite{ma00,sel00} to make analytical predictions for the
large-scale SZ effect. In a different approach, Zhang \& Pen
\cite{zha00} recently presented analytical predictions based on
hierarchical clustering and non-linear perturbation theory.

The accuracy of these different predictions are limited both by the
limited dynamical range of the simulations and the lack of detailed
comparison between the numerical and analytical methods. In this
paper, we combine the approaches of Refregier et al. \cite{ref00}
and of Cooray \cite{coo00} to carefully compare predictions from
numerical simulations to that from the extended halo model. We use
a recently developed Adaptative Mesh Refinement (AMR)
hydrodynamical code, called RAMSES (see Teyssier \cite{teyssier01}
for a complete presentation), to compute the power spectrum of the
gas pressure and of the Dark Matter (DM) density in a $\Lambda$CDM
universe. The AMR simulations provide a dynamic range of 4 orders
of magnitude and thus allow us to significantly improve upon
earlier calculations. After ensuring the proper treatment of the DM
power spectrum, the extended halo model allows us to predict the
gas-pressure power spectrum without additional free parameter. We
compare the predictions from both methods in detail, and pay
particular attention to the limitations imposed by finite mass
resolution and box size.  We then compute the resulting SZ power
spectrum and discuss the reliability and accuracy of our
prediction.

This paper is organized as follows. In \S\ref{sz}, we briefly
describe the SZ effect and derive expressions for the integrated
comptonization parameter and the SZ power spectrum.  In
\S\ref{halo}, we describe the halo model for the DM and the
baryons. In \S\ref{simulation}, we describe the numerical methods
used for the RAMSES simulations. In \S\ref{results} we present the
comparative results for the DM power spectrum, the evolution of the
density-weighted temperature, the pressure power spectrum and the
SZ power spectrum. Our conclusions are summarized in
\S\ref{conclusion}.

\section{Sunyaev--Zel'dovich Effect}
\label{sz}

The thermal SZ effect is produced from the inverse Compton
scattering of CMB photons by the hot electrons in the IGM
\cite{sun72,sun80,rep95,bir99}. The first SZ observable is the mean
comptonization parameter which can be measured from the distortion
of the CMB Planck spectrum (see \onlinecite{ste97} for a review).
Following the conventions of \cite{ref00}, we find that it is given
by
\begin{equation}
  \label{eq:y_bar}
  \overline{y} = y_{0}
  \int_0 ^{\chi_i} d\chi \overline{T}_{\rho} a^{-2}.
\end{equation}
where $\chi$ is the comoving distance, $a$ is the expansion parameter
and $y_{0}\simeq 1.710 \times 10^{-16} \left( \frac{\Omega
h^{2}}{0.05} \right)$ K$^{-1}$ Mpc$^{-1}$, for a He mass fraction of
0.24 and assuming thermal equilibrium between the ions and
electrons. The limiting distance $\chi_i$ corresponds to the
reionization redshift, for which recent theoretical studies suggest a
value around $z_i \simeq 10$ \cite{gnedin00}. The mean
density-weighted temperature of the gas is proportional to the average
pressure and is defined as
\begin{equation}
\label{eq:t_rho_def}
\overline{T}_{\rho}  \equiv
\langle \rho  T \rangle/ \langle  \rho \rangle,
\end{equation}
where $\rho$ and $T$ are the density and temperature of the gas,
respectively.

The large-scale SZ effect can also be measured from the
anisotropies that it induces on the CMB temperature.  In the
Rayleigh-Jeans (RJ) regime and in the small angle approximation,
the angular power spectrum of these fluctuations is given by (see
again \cite{ref00} for conventions)

\begin{equation}
\label{eq:cl} C_{\ell} \simeq 4 y_{0}^{2} \int_0^{\chi_i} d\chi
\overline{T}_{\rho}^{2} P_{p}\!\left(\frac{\ell}{r},\chi
\right) a^{-4} r^{-2},
\end{equation}

where $r$ is the comoving angular diameter distance, and
$P_{p}(k,\chi)$ is the 3-dimensional power spectrum of the pressure
fluctuations $\delta_{P}=(P- \langle P \rangle)/\langle P \rangle$ at
comoving distance $\chi$. In the course of our work, we found that
$\overline y$ is sensitive to the actual value of $z_i$, while the SZ
power spectrum $C_{l}$ is almost insensitive to it. For
definitiveness, we hitherto assume a fixed value for the reionization
redshift of $z_i = 10$.

\section{Halo model}
\label{halo}

\subsection{Dark Matter}
\label{halo_dm}

To model the dark matter as a collection of halos, we follow the
approach of Seljak \cite{sel00} and of Ma \& Fry \cite{ma00}.  Because
its precise predictions depend on a number of assumptions, we here
summarize the components of our halo model. The number of halos of
mass M per unit comoving volumes and per unit mass at redshift $z$ is
given by the mass function
\begin{equation}
\frac{dn}{dM} = \frac{\overline{\rho}}{M}\frac{d\nu}{dM} f(\nu),
\end{equation}
where $\overline{\rho}$ is the average matter density. The peak
height $\nu$ is defined as
\begin{equation}
\nu(M,z) \equiv \frac{\delta_{c}}{\sigma(M,z)}
\end{equation}
where $\sigma(M,z)$ is the linear {\it rms} mass fluctuation in
spheres of radius $R$ given by $M=4\pi \overline{\rho} R^{3}/3$ at
redshift $z$. This quantity can be computed from the linear power
spectrum $P_{\rm lin}(k,z)$, which we evaluate from the the BBKS
transfer function \cite{bar86} (with the conventions of
\onlinecite{pea97}) evolved with the linear growth factor. The
density threshold $\delta_{c}$ is equal to 1.68 in an Einstein-De
Sitter universe, and has a weak cosmology dependence which we
compute using the results of Kitayama
\& Suto \cite{kit96}. We adopt the standard Press-Schechter
formalism \cite{pre74}, which dictates
\begin{equation}
f(\nu) = \sqrt{\frac{2}{\pi}} e^{-\nu^{2}/2}.
\end{equation}
In order for the halos to amount to the total mass density
$\overline{\rho}$, the mass function must be normalized as
\begin{equation}
\label{eq:rho_norm}
\frac{1}{\overline{\rho}} \int_{0}^{\infty} dM \frac{dn}{dM} M =
\int_{0}^{\infty} d\nu f(\nu)=1.
\end{equation}
This is formally satisfied by the Press \& Schechter mass function,
as soon as $\sigma(M,z) \rightarrow +\infty$ for $M \rightarrow 0$.
However, in CDM-like model this divergence at small $M$'s is very
slow, thus, practically, leaving a fraction of the mass in the
background (i.e. not in collapsed halos). We account for this
background mass by adding a constant to the mass function in our
smallest mass bin (typically $10^{6} M_{\odot}$).

To each halo, we assign a Navarro, Frenk \&  White (NFW) mass
density profile \cite{nav96}, which is given by
\begin{equation}
\label{eq:nfw}
\rho(r) = \rho_{s} u(r/r_{s}), ~~u(x)=x^{-1} (1+x)^{-2}
\end{equation}
where $r$ is the comoving radius, $r_{s}$ and $\rho_{s}$ are the
characteristic radius and density, respectively . These
characteristic quantities can be written in terms of the virial
radius $r_{v}$ within which the mean density contrast is
$\delta_{v}=200$. Identifying the mass $M$ to the virial mass, we
get $M=4 \pi
\overline{\rho}
\delta_{v} r_{v}^{3}/3 = 4 \pi \int_{0}^{r_{v}} dr r^{2} \rho(r)$,
yielding $\rho_{s}=\overline{\rho} \delta_{v} c^{3} \left[ \ln(1+c)
- c/(1+c) \right]^{-1}/3$.  In the previous expression, we have
used the compactness parameter which is defined as
\begin{equation}
c(M,z) \equiv r_{v}/r_{s}.
\end{equation}
For the $\Lambda$ CDM model we consider here, we adopt the
functional form of $c$ given by \cite{coo+00,coo00}
\begin{equation}
\label{eq:c_mz}
c(M,z) \simeq g(z) \left[
\frac{M}{M_{*}(z)} \right]^{-h(z)},
\end{equation}
where $g(z) \simeq 10.3(1+z)^{-0.3}$ and $h(z) \simeq
0.24(1+z)^{-0.3}$.  Here, $M_{*}(z)$ is the non-linear mass scale
defined by $\nu(M_{*},z) \equiv 1$. The bias for the clustering of
the halos is given by the Mo \& White formalism \cite{mo96},
namely by
\begin{equation}
\label{eq:b_nu}
b(M,z) = 1 + \frac{\nu^{2}-1}{\delta_c}.
\end{equation}

Assuming that the matter distribution  can be modeled as a
collection of these halos, the matter power spectrum is given by
\cite{sel00,ma00}
\begin{equation}
\label{eq:p_rhodm}
P(k,z) = P_{1}(k,z) + P_{2}(k,z),
\end{equation}
where the  terms   correspond  the   1-halo  (Poisson)   and 2-halo
(clustering) contribution. They are given by
\begin{equation}
P_{1}(k)= \int_{0}^{\infty} dM ~\frac{dn}{dM} \left[
  \frac{\tilde{\rho}(k,M)}{\overline{\rho}} \right]^{2}
\end{equation}
and
\begin{equation}
P_{2}(k)= \left[ \int_{0}^{\infty} dM ~\frac{dn}{dM} b(M)
  \frac{\tilde{\rho}(k,M)}{\overline{\rho}} \right]^{2} P_{\rm lin}(k)
\end{equation}
where the radial Fourier transform of the density profile is given
by
\begin{equation}
\label{eq:rft}
\tilde{\rho}(k,M) = 4 \pi \int_{0}^{r_{v}} dr~r^2 \rho(r,M) \frac{\sin kr}{kr}.
\end{equation}
Conveniently, $\tilde{\rho}(0)=M$. As noted by Seljak \cite{sel00},
the fact that on large scales $P(k)$ must approach $P_{lin}(k)$
imposes the non-trivial constraint
\begin{equation}
\frac{1}{\overline{\rho}} \int_{0}^{\infty} dM~\frac{dn}{dM} M b(M) =
\int_{0}^{\infty} d\nu~f(\nu) b(\nu) = 1,
\end{equation}
which formally holds in the case $\sigma(M,z) \rightarrow +\infty$
for $M \rightarrow 0$. As for the mass function, we practically
force this normalization to be numerically exact by adding a
constant to $b(M)$ in our smallest-mass bin.

\begin{figure*}[httb]
\centerline{\epsfig{file=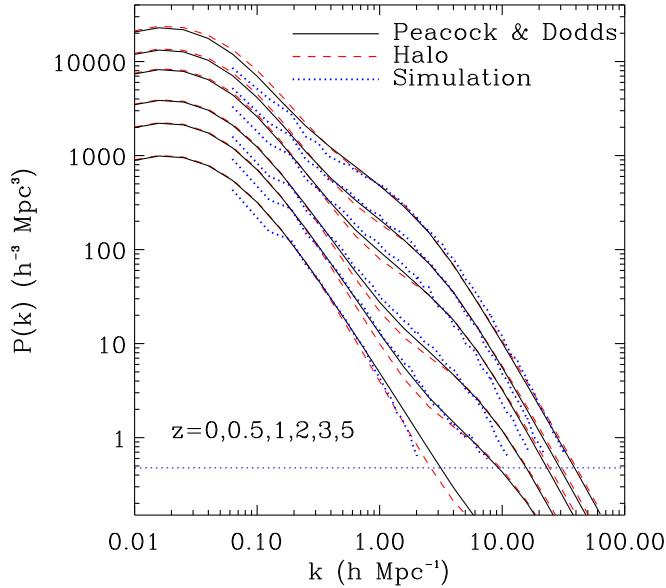,width=90mm}}
\caption{
Power spectrum of the DM density at different redshifts for the
$\Lambda$CDM model. The solid, dashed and dotted curves correspond
to the Halo model, the Peacock \& Dodds fit, and the simulations,
respectively. The thin horizontal dotted line corresponds to the
Poisson noise in the simulations. }
\label{fig:p_rhodm}
\end{figure*}

The resulting power spectra at different redshifts are shown on
Figure \ref{fig:p_rhodm} for the $\Lambda$CDM model (with
parameters listed in
\S\ref{sim_param}). Also shown are the power spectra from the
Peacock \& Dodds \cite{pea96} fitting formula. As noted by
\cite{coo+00}, the agreement is good for all redshifts. Note that
the only tuning involved is that of the functional form of the
compactness parameter $c(M,z)$ (Eq.~[\ref{eq:c_mz}]).

\subsection{Virial Temperature}
To model the baryons, we first consider the virial temperature of a
halo which is given by
\begin{equation}
\label{eq:t_vir}
kT_{v} = \frac{G\mu m_{p} M}{2 \beta_{v} r_{v,{\rm phys}}}
 \simeq  7.82 (1+z) \beta_{v}^{-1}
 \Omega_{m}^{\frac{1}{3}} \left(\frac{M}{10^{15} h^{-1} M_{\odot}}
 \right)^{\frac{2}{3}} \left( \frac{\mu}{0.59}
 \right) \left( \frac{\Delta_{c}}{178} \right)^{\frac{1}{3}} {\rm
 keV},
\end{equation}
where $r_{v,{\rm phys}}$ is the virial radius in physical
coordinates and $\Delta_{c}(z,\Omega_{m},\Omega_{\Lambda})$ is the
average virial overdensity which can be evaluated using the fitting
formulae of \cite{kit96}. The value $\mu=0.59$ (the number of
particles per proton mass) corresponds to a hydrogen mass fraction
of 76\%. The factor $\beta_{v}$ is exactly equal to 1 for a
truncated singular isothermal sphere. In the general case, it can
be considered as an unknown normalization parameter, whose exact
value is determined using numerical simulations. Values of
$\beta_{v}$ between $1$ and $1.2$ provide good fits to numerical
simulations \cite{eke96,bryan98}. Here, we adopt $\beta_{v}=1$.

\begin{figure}
\centerline{\epsfig{file=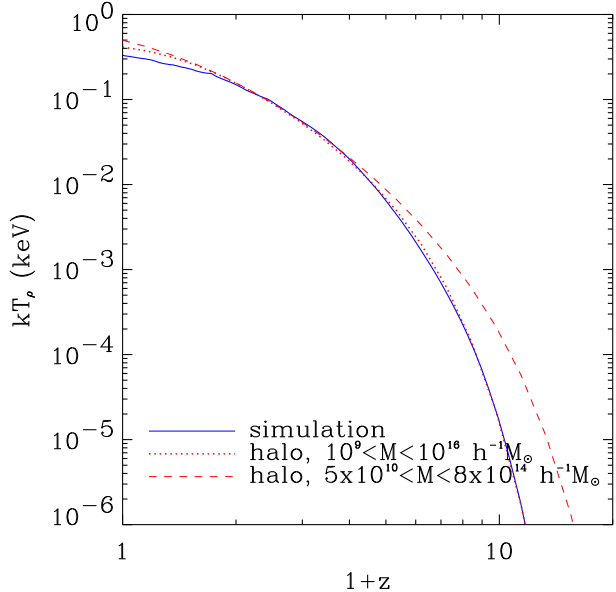,width=90mm}}
\caption{Evolution of the density-weighted temperature
$\overline{T}_{\rho}$ of the gas. The prediction from the
simulations is shown as the solid line. The prediction for the halo
model is shown for different mass ranges $M_{\rm min}<M<M_{\rm
max}$ as dashed and dotted lines.}
\label{fig:tz}
\end{figure}

Taking the gas temperature to be equal to the virial temperature,
we can compute the density-weighted temperature
(Eq.~[\ref{eq:t_rho_def}]) of the gas which is given by
\begin{equation}
\label{eq:t_rho}
\overline{T}_{\rho} = \frac{1}{\overline{\rho}}
  \int_{M_{\rm min}}^{M_{\rm max}} dM~\frac{dn}{dM} M T_{v},
\end{equation}
where $M_{\rm min}$ and $M_{\rm max}$ are mass limits introduced to
model the finite mass resolution and box size of our numerical
simulation (see \S\ref{sim_resol}). Note that, to ensure proper
normalization of the total DM density (Eq.~[\ref{eq:rho_norm}]),
this mass range is not used to compute the power spectrum of the DM
density (Eq.~[\ref{eq:p_rhodm}]). The resulting redshift evolution
of this temperature is plotted on Figure~\ref{fig:tz}, for
different mass ranges.

\subsection{Gas Profile}
\label{gasprof}
To derive the profile of the gas pressure $p_{g}$ of halos, we
further assume that the gas is in hydrostatic equilibrium and thus
obeys
\begin{equation}
\label{eq:hydrostat}
\frac{d p_{g}}{d r} = - \frac{G \rho_{g}(r) M(<r)}{r^{2}}
\end{equation}
where $\rho_{g}$ is the gas density and $M(<r)$ is the total mass
within radius $r$. Approximating the total mass as that of the Dark
Matter only, it is easy to show that, for the NFW profile
(Eq.~[\ref{eq:nfw}]),
\begin{equation}
\label{eq:mr_nfw}
M(<r) = 4 \pi \rho_{s} r_{s}^{3} \left[ \ln (1+\frac{r}{r_{s}})
- \frac{r}{r+r_{s}} \right]
\end{equation}

We need further assumptions to integrate
Equation~(\ref{eq:hydrostat}) for the pressure profile $p_{g}(r)$.
As Cooray \cite{coo00}, we consider the simplest assumption, namely
that the gas is ideal ($p_{g}=\frac{k_{B}}{\mu m_{p}} \rho_{g}
T_{g}$) and isothermal within each halo, with a temperature equal
to the virial temperature, i.e. $T_{g}(r)=T_{v}$. In this case, the
hydrostatic equilibrium equation for the NFW mass profile yields
\begin{equation}
\rho_{g}(r) = \rho_{g0} v(r/r_{s},\lambda),~~~
v(x,\lambda) = e^{-\lambda}(1+x)^{\frac{\lambda}{x}},
\end{equation}
with
\begin{equation}
\lambda = \frac{4 \pi \mu m_{p} G \rho_{s} r_{s}^{2}}{k_{B}T_{v}} .
\end{equation}
This agrees with \cite{coo00} up to a different power for $r_{s}$
in the last expression. The normalization $\rho_{g0}$ of the gas
density profile is set by requiring that the baryon mass fraction
is equal to that of the universe, i.e. by
\begin{equation}
M_{g} = 4 \pi \int_{0}^{r_{v}} dr~r^{2} \rho_{g}(r)
= \frac{\Omega_{b}}{\Omega_{m}} M.
\end{equation}
Following \cite{coo00}, we avoid an abrupt cutoff at the virial
radius by multiplying the profile function $v(x)$ by an apodizing
function which we choose to be $\left| E\left(\frac{x-c}{\sqrt{2}}
\right)\right|$, where $E(x)$ is the error function.

\begin{figure}
\centerline{\epsfig{file=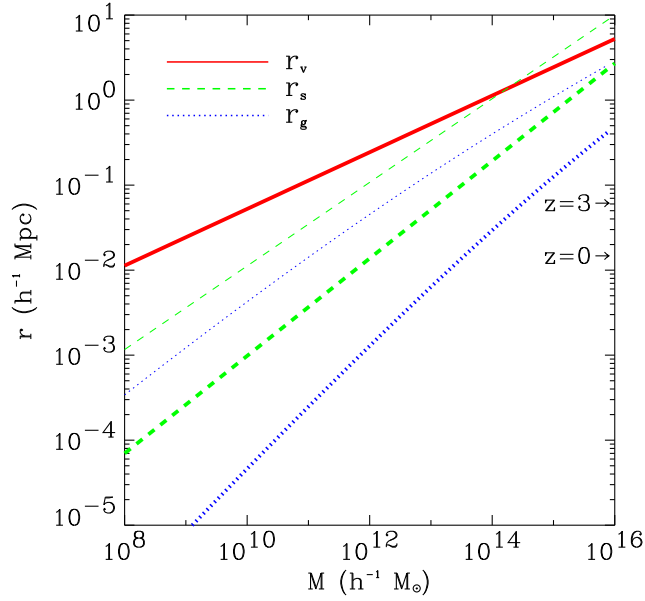,width=90mm}}
\caption{Characteristic comoving radii for the halo model as a function of halo
mass. The virial radius $r_{v}$ (solid line), the NFW scale radius
$r_{s}$ (dashed lines) and the gas inner radius $r_{g}$ (dotted
lines) are all shown at $z=0$ (thick lines) and $z=3$ (thin lines).
Note that the curves for the virial radius are identical for all
redshifts. It is the interplay between these radii which determine
the behavior of the gas pressure power spectrum on intermediate
scales. We indicate by arrows the comoving spatial resolution of
the numerical simulation at each redshift.}
\label{fig:rs}
\end{figure}

In the limit of small radii, the gas density approaches
$\rho_{g}(r)
\propto e^{-\frac{r}{r_{g}}}$, where $r_{g}=\frac{2r_{s}}{\lambda}$ is
the characteristic radius where the gas central density approaches
a constant. It can be considered as a ``core radius'' in the gas
distribution. The three characteristic radii $r_{v}$, $r_{s}$ and
$r_{g}$ are plotted as a function of mass $M$ at $z=0$ and $z=3$ on
figure~\ref{fig:rs}. The inner radius $r_g$ scales roughly as $r_s$
with mass, but tends to increase as $z$ increases. This is due to
the fact that, for a given mass, halos tend to have a higher
temperature at high redshift (Eq. \ref{eq:t_vir}). In this model,
the gas distribution is thus predicted to be more peaked at low redshift.

\subsection{Pressure Power Spectrum}

\begin{figure}
\centerline{\epsfig{file=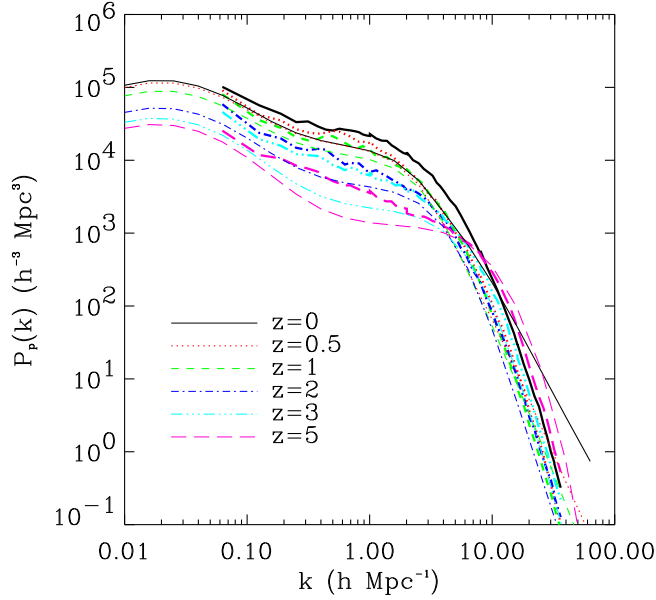,width=90mm}}
\caption{Power spectrum of the gas pressure at different redshifts.
The thin smooth curves correspond to the analytical halo model
(with $M_{\rm min}=5 \times 10^{10} h^{-1} M_{\odot}$ and $M_{\rm
max}=8\times10^{14} h^{-1} M_{\odot}$), while the thick broken
lines correspond to the simulations.}
\label{fig:pps}
\end{figure}

As for the dark matter density (Eq.~[\ref{eq:p_rhodm}]), the
pressure power spectrum $P_{p}(k)$ is related to the pressure
profile by
\begin{equation}
\label{eq:p_pressure}
P_{p}(k,z) = P_{p1}(k,z) + P_{p2}(k,z),
\end{equation}
where 1-halo and  2-halo terms are given by
\begin{equation}
P_{p1}(k)= \int_{M_{\rm min}}^{M_{\rm max}} dM ~\frac{dn}{dM}
\left[
  \frac{\tilde{p}_{g}(k,M)}{\overline{p}_{g}} \right]^{2}
\end{equation}
and
\begin{equation}
P_{p2}(k)= \left[ \int_{M_{\rm min}}^{M_{\rm max}} dM
~\frac{dn}{dM} b(M)
  \frac{\tilde{p}_{g}(k,M)}{\overline{p}_g} \right]^{2} P_{\rm lin}(k),
\end{equation}
where ($M_{\rm min}$, $M_{\rm max}$) is the mass range introduced
in Equation~(\ref{eq:t_rho}). Here, $\tilde{p}(k,M)$ is the radial
Fourier transform (Eq.~[\ref{eq:rft}]) of the gas pressure profile
$p_{g}(r)$. The mean gas pressure is
$\overline{p}_{g}=\frac{k_{B}}{\mu m_{p}}
\overline{\rho}_{g}
\overline{T}_{\rho}$, where the mean gas density is assumed to be
$\overline{\rho}_{g} = \frac{\Omega_{b}}{\Omega_{m}}
\overline{\rho}$. The pressure power spectrum is plotted at different
redshifts on Figure~\ref{fig:pps}.

\begin{figure}
\centerline{\epsfig{file=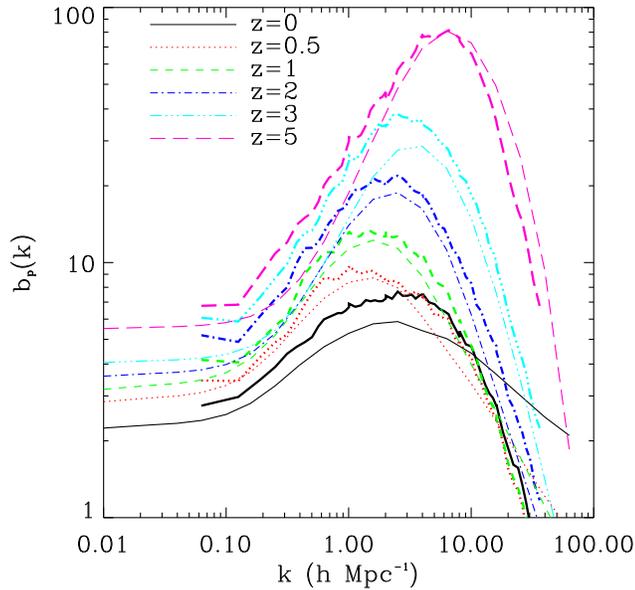,width=90mm}}
\caption{Bias $b_{p}(k)$ of the gas pressure with respect to the DM
density (see text) at different redshifts. As before, predictions
for both the halo model (with $M_{\rm min}=5 \times 10^{10} h^{-1}
M_{\odot}$ and $M_{\rm max}=8\times10^{14} h^{-1} M_{\odot}$) and
for the simulations are shown.}
\label{fig:bps}
\end{figure}

It is useful to consider the bias of the gas pressure relative to
the dark matter density,
\begin{equation}
\label{eq:bp}
b_{p}^{2}(k,z) \equiv \frac{P_{p}(k,z)}{P(k,z)}.
\end{equation}
It is shown for different redshifts on Figure~\ref{fig:bps}.

\begin{figure}
\centerline{\epsfig{file=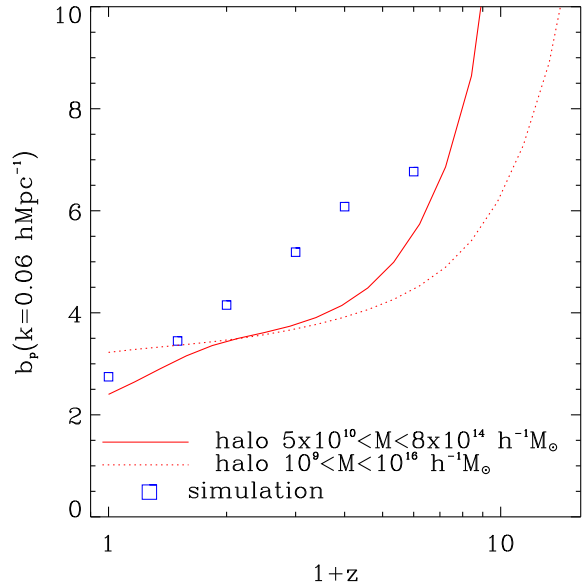,width=90mm}}
\caption{Large scale bias $b_{p}$ of the gas pressure as a function of
redshift. The predictions from the halo model (for different mass
ranges) and simulation are shown for $k=0.06 h^{-1}$ Mpc.}
\label{fig:bp_z}
\end{figure}

On large scales ($k\rightarrow 0$), the 2-halo term dominates and
$P(k)\simeq P_{\rm lin}(k)$, so the pressure bias reduces to
\begin{equation}
\label{eq:bp0}
b_{p}(k\rightarrow 0) \simeq
\frac{1}{\overline{\rho}\overline{T}_{\rho}}
\int_{M_{\rm min}}^{M_{\rm max}} dM~\frac{dn}{dM} b(M) MT.
\end{equation}
In this limit, the pressure bias is thus simply the
pressure-weighted average of the halo bias $b(M)$. The large scale
pressure bias (at $k=0.06 h$ Mpc$^{-1}$ derived from
Eq~[\ref{eq:bp}]) is plotted in Figure~\ref{fig:bp_z} for different
mass ranges ($M_{\rm min}$, $M_{\rm max}$).

\section{Simulations}
\label{simulation}

The halo model we described above relies on strong assumptions,
namely that the mass density and pressure can be described
statistically as a collection of spherically symmetric, isothermal
halos in hydrostatic equilibrium. Although complicated structures,
like walls and filaments, are known to exist in the hierarchical
clustering picture, it remains to be established whether their
influence on the resulting pressure power spectrum is sufficient to
invalidate our simple halo model.  In the following, we describe
the high-resolution hydrodynamical simulations we have used to
check the domain of validity of our analytical approach.

\subsection{Cosmological simulations using Adaptive Mesh Refinement}
\label{amr}
Adaptive Mesh Refinement (AMR) is now widely used in astrophysics
\cite{klein94,bryan97,truelove98,khokhlov99,ricker99} and offers an
interesting alternative to the Smooth Particle Hydrodynamics (SPH)
algorithm. AMR was recently adapted to the cosmological framework
by several groups \cite{bryan97,kravtsov97}. The main advantage of
this grid-based method, compared to  the well-known Smooth Particle
Hydrodynamics (SPH) method \cite{monaghan92}, is its ability to
capture shock-waves and contact discontinuities using only a few
resolution elements (typically 2-3 cells). In comparison, SPH
demands at least 50 particles per resolution element, in order to
minimize Poisson noise related to the discrete nature of the
algorithm \cite{kang94}.

The basic AMR idea is to refine recursively an initially coarse grid
(defined here as the coarse level $\ell=0$) with smaller and smaller
cells of comoving size $\Delta x = 2^{-\ell}$ ($\ell > 0$).  These
refinements take place in regions where interesting features of the
flow demand better spatial resolution (see
Figure~\ref{fig:cluster}). For example, in fluid dynamics, one usually
refines sharp discontinuities like shocks or contact surfaces.  In
cosmology, refinements are used within high-density regions, in order
to resolve properly the gravitational collapse of small scale halos
and the progressive build-up of large, massive ones. Due to the highly
compressible nature of the self-gravitating cosmological fluid, this
property improves dramatically the quality of the numerical solution.

In this paper, we use the results of a numerical simulation
obtained using the newly developed RAMSES code, described in
details in \cite{teyssier01}. This code is based on general AMR
techniques, with several specific properties that we briefly recall
here. Contrary to patch-based AMR, where a small number of
individual grids (say a few thousands) define rectangular regions
of higher and higher resolution (see for example
\cite{bryan97,truelove98,ricker99}), RAMSES is a cell-based AMR:
each cell can be independently divided in sub-cells and thus can
describe complex flow features of arbitrary geometry (see for
example \cite{kravtsov97,khokhlov98}). The price to pay for this
improved flexibility is a higher computational complexity: to
describe the AMR grid, we use the ``Fully Threaded Tree'' data
structure described by Khokhlov \cite{khokhlov98}, where cells are
placed in a recursive tree connecting each cell to its 8 sister
cells and its 6 neighboring cells, with minimum memory overhead.

The N-body module of RAMSES is similar to the Adaptive Refinement
Tree (ART code) of Kratsov, Klypin \& Khokhlov \cite{kravtsov97},
with some differences outlined in Teyssier \cite{teyssier01}.  The
hydrodynamics module of RAMSES uses an unsplit, second-order
Godunov scheme. Time integration proceeds in two different ways in
RAMSES, either using a single, small time step for all levels of
refinement or using nested, adaptive time steps in order to speed
up the calculation for coarse levels. The last option was retained
for cosmological simulations and turned out to speed up the
calculation by a factor of 10, compared to the single-time step
case. A complete presentation of the performances obtained by
RAMSES is presented in Teyssier \cite{teyssier01}, using standard
tests to validate our algorithm and its application to cosmological
hydrodynamics.

\subsection{Simulation Parameters}
\label{sim_param}

In this paper, we present the results of a RAMSES simulation for a
flat $\Lambda$CDM model with $\Omega_{M} = 0.3$, $\Omega_{\Lambda} =
0.7$, $\Omega_{B}=0.039$. We use the transfer function given in
\cite{ma98}. The resulting power spectrum is normalized to the COBE
data \cite{white95} (this corresponds in our case to
$\sigma_{8}=0.93$). The comoving box size was set to 100$h^{-1}$Mpc,
imposing a maximum scale length in the simulation: $k_{min}=0.06$ $h$
Mpc$^{-1}$. Initial conditions are specified using an initial grid of
$128^3$ dark matter particles. This corresponds to a mass resolution
of $M_{\rm min}=4 \times 10^{10}~M_{\odot}$. The particles were
initially displaced according to the Zel'dovich approximation up to
the starting redshift $z_i \simeq 55$. The baryon density and velocity
fields were perturbed accordingly. We assume that baryons are
described by a purely adiabatic, $\gamma = 5/3$, fully ionized
plasma. Since we are studying the SZ effect, we are mainly concerned
by the pressure field of the baryons, which, at the scales of interest
here, should remain unaffected (at least to first order) by neglecting
others physical processes such as cooling, star formation and
supernovae energy feedback. We plan to study the influence of these
others physical inputs in a future paper.

The key parameters of any AMR simulation are the refinement
criteria used to dynamically create the refinement tree. We use the
``quasi-Lagrangian'' approach described in Kratsov, Klypin \&
Khokhlov \cite{kravtsov97}. The idea is to mark for refinement any
cells whose gas {\it or} dark matter overdensity exceeds a
level-dependent threshold given by
\begin{equation}
\frac{\rho}{\bar \rho} (\ell=0,5) = 1, 80, 640, 5120, 20480, 163840
\end{equation}
The coarse grid ($\ell=0$) is nothing but the $128^3$ particle grid
used to set up the initial conditions. The $\ell=0$ refinement
criteria ($\rho/{\bar \rho} > 1$) ensures that, initially, the
whole computational volume is covered by a $256^3$ grid ($\ell=1$).
In this way, initial small scale perturbations are sampled by two
grid points, which turns out to be necessary to avoid small scale
power damping. At later times, large voids appear in the simulation
where the resolution is locally degraded down to the $128^3$ coarse
grid. For higher levels of refinements, the density thresholds are
chosen in order to refine cells that contains between 5-10
particles (or fluid mass elements). This ``quasi-Lagrangian''
approach preserves the initial resolution in {\it physical}
coordinates, adapting the local {\it comoving} spatial resolution
to collapsing mass elements (see \cite{knebe00} for a complete
discussion).

The maximum level of refinement reached at the end of the present
run is $\ell=6$, which gives a formal spatial resolution of
$8192^3$ (or 12 kpc $h^{-1}$ comoving) at $z=0$. The comoving
spatial resolution scales roughly with redshift as $(1+z)^{-1}$. We
also use the adaptive time-stepping procedure of RAMSES to speed up
the computation: 157 time steps only were necessary at the coarse
level while 8185 time steps were necessary at the finest level. The
Layzer-Irvine energy conservation was better than $1\%$ at the end
of the simulation. Starting with $19 \times 10^6$ cells at
$z_i=55$, we end up with $28 \times 10^6$ cells at $z=0$ in the
tree structure (including the coarse level). Six different output
times were considered: $z=0,0.5,1,2,3$ and $5$.

\subsection{Effect of finite mass resolution and box size}
\label{sim_resol}

The question that arises when interpreting numerical simulations is
that of its range of validity. Here, we discuss three limitations
of cosmological simulations which can introduce spurious effects:
the finite box size, the mass resolution and the spatial
resolution.

Since we solve for the gravitational evolution of the system
assuming periodic boundary conditions, the power on scales larger
than the box size is completely suppressed. This translates into a
maximum mass above which no halo can form. Indeed, the probability
to find a large halo of mass, say, $10^{16} h^{-1} M_{\odot}$ in
our computational box is close to zero. An estimate for the maximum
mass $M_{\rm max}$ which corresponds to our box size is given by
\begin{equation}
n(>M_{\rm max})L_{\rm box}^3 = 1
\end{equation}
where $n(>M)$ is the cumulative Press-Schechter mass function. For
$L_{\rm box}$ = 100 $h^{-1}$Mpc, this gives $M_{\rm max} \simeq 7
\times 10^{14} h^{-1} M_{\odot}$. Note that, on large
scales, we are dominated by the ``cosmic variance'' of the
simulation: depending on the exact random field realization for the
initial conditions, we may still form by chance a very large
cluster in the simulation. The next step is to check the
convergence of various quantities with respect to $M_{\rm max}$,
using our analytical model. As we show below (\S\ref{results_pp}),
we find that good convergence in the SZ power spectrum is achieved
for $M_{\rm max}=2 \times 10^{15} h^{-1} M_{\odot}$, or
equivalently $L_{\rm box}$ = 400 $h^{-1}$Mpc. On the other hand, for a
fixed number of particles, increasing the box size inevitably
degrades the mass and spatial resolutions of the simulation. Our
current choice of $L_{\rm box}$ = 100 $h^{-1}$Mpc is a trade off
between these two opposite constraints.

The mass resolution, $M_{\rm min} \propto k_{\rm max}^{-3}$, is
related to the small scale power in the initial conditions. No
object with $M \le M_{\rm min}$ will form, leaving instead a cold,
smooth background between collapsing objects.
In order to account for these two effects (finite box size and mass
resolution), we introduced in the analytical description a mass
range where the different integrals are computed. This limited mass
range is supposed to mimic the lack of large mass haloes due to
the finite box size ($M_{\rm max}$), and the lack of small mass
halos due to the power cut off at small scale ($M_{\rm min}$).

The final numerical effect which needs to be examined is the
spatial resolution. The spatial resolution is related to the
minimum scale below which the code is not able to solve the
equations. For example, in a standard Particle Mesh code, this
scale is equal to 2 cell size of the underlying Cartesian mesh,
below which the force between 2 interacting particles goes smoothly
to zero. In practice, this acceleration cut off at small radii
results in a damping of the power spectrum up to scales as large as
8 comoving cell size. For gas dynamics, the problem of spatial
resolution is even more crucial. Shock waves need at least 2 cells
to be properly resolved. If small scale velocity fluctuations are
not properly sampled, no shock dissipation can occur {\it at all}.
In this case, the small scale density fluctuations are only
compressed adiabatically, i.e. they remain cold, even though their
temperature should rise to the virial temperature. This can affect
both the mean and the variance of the computed gas pressure.

During the course of gravitational collapse, the initial
perturbations contract at smaller and smaller comoving scales,
making the spatial resolution issue even more crucial. The spatial
resolution therefore translates directly into a minimum mass, below
which halo collapse cannot be resolved properly by the code. For a
Particle-Mesh code, coupled to a standard gas dynamics code, the
resulting minimum mass can be as high as 100 particles, for a halo
to be properly resolved. Most of the initial small scale power is
then lost this way. As discussed in section \S\ref{amr}, AMR
techniques allows one to dynamically adapt the comoving resolution to
the collapsing mass element. In light of the results obtained in
this paper, we claim that the RAMSES code allows us to resolve
properly small scale shock heating, down to the minimal halo mass
imposed by the initial conditions.

\subsection{Computation of the Power spectrum}
In this paper, we need to compute the DM density power spectrum,
together with that of the gas pressure. Since the spatial
resolution goes from 96 kpc $h^{-1}$ at $z=5$ down to 12 kpc
$h^{-1}$ at $z=0$ for a box size of 100 Mpc $h^{-1}$, the dynamical
range imposes a severe challenge in computing power spectra. The
brute force approach would consist in Fourier analyzing each field
on a large Cartesian grid of $1024^3$ for $z=5$ (barely feasible on
a supercomputer), up to a Cartesian grid of $8192^3$ for $z=0$
(completely out of reach).

A more appropriate approach is described in Jenkins et al.
\cite{jenkins98} and Kravtsov \& Kylpin \cite{kravtsov99}. The idea is
to define each field (dark matter density or gas pressure) in a nested
structure of tractable mesh (we use $128^3$ cells). At each level
$\ell$, we divide the whole computational volume in $\ell^3$ Cartesian
$128^3$ sub-cubes and add up the fields computed in all these
sub-cubes together. We then use a standard Fourier power spectrum
estimation for the resulting field. Each scale $\ell$ provides a
reliable power spectrum estimation between $2^{\ell} \times [k_{\rm
min}, k_{\rm max}]$, where $k_{\rm min} \approx 8(2\pi / L_{\rm box})$
and $k_{\rm max}\approx 16(2\pi / L_{\rm box})$. These limits are
determined empirically, by comparing the results obtained for 3 nested
$128^3$ grids and for a single $512^3$ grid. The resulting power
spectrum estimation is obtained by combining the different ``band
power'' estimates together. Finally, we subtract from the dark matter
density power spectrum the shot noise (or Poisson noise) due to the
discrete nature of the particle distribution.

\section{Results}
\label{results}

\subsection{Dark Matter Power Spectrum}
Figure~\ref{fig:p_rhodm} shows the dark matter power spectrum
$P(k)$ at different redshifts, derived from the Peacock \& Dodds
\cite{pea96} formalism, the halo method and the simulations. The
finite box size and the mass resolution (see the Poisson noise
limit on the figure) restricts predictions of the simulations to
the range $0.06 \lesssim k \lesssim 60$ $h$ Mpc$^{-1}$. For all
redshifts, the simulations agree very well with Peacock \& Dodds
within this range, which spans 3 orders of magnitude. This range is
consistent with the formal spatial resolution of the simulation,
which spans 4 order of magnitude, after accounting for the small
scale power damping discussed in \S\ref{sim_resol}. In comparison,
the simulations used by Refregier et al. \cite{ref00} and Seljak,
Burwell, \& Pen \cite{sel99} agreed with Peacock \& Dodds only
within one order of magnitude in $k$, and therefore seriously
limited the SZ predictions of these authors \cite{ref00}. The
present simulations therefore allow us to significantly improve
over these previous works.

As stated in \S\ref{halo_dm}, the halo model prediction for $P(k)$
is in good agreement with the Peacock \& Dodds spectrum for all
redshifts (see \cite{coo+00}). Small deviations of the order of a
few percents can be noticed for $k \lesssim 10$ $h$ Mpc$^{-1}$, but
are acceptable for our purposes. Note in particular that the
evolution of the break in the power spectrum between the linear and
the nonlinear regime is well modeled by the halo model.

\subsection{Temperature Evolution}
Figure~\ref{fig:tz} shows the evolution of the mass-weighted
temperature $\overline{T}_{\rho}(z)$ (Eq.~[\ref{eq:t_rho_def}]) as
predicted from the halo model (Eq.~[\ref{eq:t_rho}]) and from the
simulation. For the halo model, different mass cut off ranges $M_{\rm
min}<M<M_{\rm max}$ (see Eq.~[\ref{eq:t_rho}]) are examined.

The halo prediction is sensitive to $M_{\rm max}$ and $M_{\rm min}$ at
low and high redshift, respectively. At high redshift, the
characteristic non-linear mass scale $M_{\star}$ is close to the mass
resolution of the simulation, for which no shock heating can
occur. Logically, this translates in an underestimation of the mean
temperature. At low redshift, the non-linear mass scale $M_{\star}$
approaches dangerously the maximum mass scale due to the finite box
size. Here again, the temperature is underestimated, but this time, it
is due to the absence of very high mass halos.

A good overall agreement (mean temperature and power spectrum
evolution) between the simulation and the analytical model is
obtained for the following mass range:
\begin{equation}
\label{eq:m_fiduc}
5 \times 10^{10} < M < 8 \times 10^{14} h^{-1} M_{\odot}
\end{equation}
It is reassuring that this best-match mass limits are close to the
natural limits imposed by the simulation, as shown in
\S\ref{sim_resol}.

For a mass range of $5\times 10^{10} < M < 4 \times 10^{14} h^{-1}
M_{\odot}$, the mean temperature evolution for the halo model (not
shown on Figure~\ref{fig:tz}) is virtually indistinguishable from
the simulation curve, with $\overline{T}(z=0) \simeq 0.34$ and 0.33
keV, respectively. However, as we will see below
(\S\ref{results_pp}), the agreement with the pressure power
spectrum slightly worsens. For $M_{\rm max}= 8 \times 10^{14}$ (our
fiducial limit) and $M_{\rm max} > 10^{16}$ $h^{-1} M_{\odot}$ (the
converged asymptotic limit) the halo model predicts
$\overline{T}(z=0) \simeq$ 0.41 and 0.50 keV, respectively.

\subsection{Mean Comptonization Parameter}
Our calculation of the evolution of the density-weighted
temperature allows us to compute the mean comptonization parameter
(see Eq.~[\ref{eq:y_bar}]). For the simulation, we find
$\overline{y}=2.65\times10^{-6}$. The halo model with our fiducial
mass limits (Eq.~[\ref{eq:m_fiduc}]) predicts a value of
$\overline{y}=2.84\times10^{-6}$. As expected, the agreement is
slightly better for $M_{\rm max} \simeq 4 \times 10^{14} h^{-1}
M_{\odot}$ with $\overline{y}=2.72 \times10^{-6}$. In all cases,
these values are similar to those found in other studies (eg.
\cite{das99,ref00,sel00}). It is interesting to study how our
prediction for $\overline{y}$ depends on the mass limits.  We find
experimentally that
\begin{equation}
\overline{y} \simeq 2.84\times 10^{-6}
\left( \frac{M_{\rm min}}{5\times 10^{10} h^{-1} M_{\odot}} \right)^{-0.06}
\left( \frac{M_{\rm max}}{8\times 10^{14} h^{-1} M_{\odot}}
  \right)^{0.01}.
\end{equation}
Since length scales scale as $M^{\frac{1}{3}}$, this means that
$\overline{y} \propto \Delta^{-0.2} L^{0.03}$, where $\Delta$ is the
effective resolution length and $L$ is the box size of the
simulation. This quantity is thus quite sensitive to the mass
resolution, but not very sensitive to the box size. This should be
kept in mind when interpreting and comparing the predictions of
simulations. The asymptotic value corresponding, effectively, to an
infinite mass range is $\overline{y} \simeq 3.01\times 10^{-6}$ (for
$z_{i}=10$, see \ref{sz}).

\subsection{Projected Map of the Pressure}
Figure~\ref{fig:mapy} shows a map of the $y$-parameter projected
through one side of the $100 h^{-1}$ Mpc box at $z=0$. Clusters
produces fluctuations in $y$ of the order of $10^{-4}$--$10^{-6}$,
while supercluster filaments produce fluctuations of the order of
$10^{-6}$--$10^{-8}$ . This figure is quite similar to the
corresponding figure from Refregier et al. \cite{ref00} derived
from the Moving Mesh Hydrodynamical code (MMH) of Pen
\cite{pen95,pen98}. Note however, the more circular appearance of
clusters in our simulations, compared to the more elongated cores
seen in that of Refregier et al. \cite{ref00}. The number of small
mass objects is also much higher in our simulation. The main
difference between RAMSES and MMH is the spatial resolution. The
improved dynamical range of RAMSES compared to MMH yields to the
fragmentation of large scale filements into small clumps. The
effective mass resolution in RAMSES is consequently much smaller
than the one reached with MMH. Moreover, substructures within large
halos, clearly visible in Figure~\ref{fig:mapy}, are able to
survive to tidal stripping much longer in RAMSES than in MMH. This
explains the more elongated cores of large halos in MMH
simulations.

\begin{figure}
\vspace*{12.cm}
\caption{Map of the $y$-parameter projected through on face of the
$100 h^{-1}$ Mpc box at $z=0$.}
\label{fig:mapy}
\end{figure}

Figure~\ref{fig:cluster} shows a close up of one the clusters in
the AMR simulation. Panel (a) shows the map of the $y$-parameter
for the cluster, while the panel (b) shows the AMR grid for this
region. Notice how the AMR grid has adapted to provide higher
resolution for the cluster core and for the two substructures below
the cluster. The virial radius for this cluster is about $1$-$1.5
h^{-1}$ Mpc, and thus almost entirely fills the displayed region
($3.12 h^{-1}$ Mpc on a side). This figure also shows that the
structure of individual clusters is more complicated than that
assumed in our halo model. In the following, we show that the halo
model nevertheless provides an acceptable prediction for the
pressure power spectrum.

\subsection{Pressure Power Spectrum}
\label{results_pp}
Figure~\ref{fig:pps} shows the power spectrum of the gas pressure
at different redshifts for both the fiducial halo model (with our
fiducial mass range of Eq.~[\ref{eq:m_fiduc}]) and for the
simulations. The overall agreement is remarkable: the shape of the
power spectra agree approximately for both methods, along with the
scaling with redshifts which changes order above and below $k\sim 3
h^{-1}$ Mpc$^{-1}$. According to both method, the pressure power
spectrum does not evolve as much as the DM density power spectrum.
Qualitatively, this is due to the fact that, at large redshift the
smaller amplitude of the density power spectrum is compensated by
the larger biasing of peaks with high temperature which dominate
the pressure power spectrum.

There are nevertheless quantitative differences: the halo model
predictions have somewhat different normalization at low $k$ and
not exactly the same shape at large $k$. This can be seen more
clearly by studying Figure~\ref{fig:bps}, which shows the bias
$b_{p}(k)$ of the gas pressure (Eq.~[\ref{eq:bp}]) at different
redshifts. Again both the simulation and fiducial halo model are
displayed. Note that we used the halo-model power spectrum in the
denominator in both cases, so that Figure~\ref{fig:bps} is nothing
but a change of scaling of the pressure power spectrum.

\begin{figure}
\vspace*{0.2cm}
\vspace*{7.0cm}
\vspace*{0.2cm}
\caption{Close up on a cluster in the simulation.
(a) Map of the $y$-parameter for the cluster. The logarithmic color
scale ranges from $y=10^{-7}$ to $10^{-4}$.
The image size is $3.12 h^{-1}$Mpc
on a side. (b) AMR grid for this region with levels from $\ell=1$ to 6.}
\label{fig:cluster}
\end{figure}

The difference in shape at large $k$ is more pronounced at low
redshifts where the simulation pressure spectrum falls off faster than
the halo model. The shape of the pressure power spectrum at
intermediate and small scales is determined by the interplay between
the different characteristic radii $r_{v}$, $r_{s}$ and $r_{g}$ (see
\S\ref{gasprof}).  Figure~\ref{fig:rs} shows that at low redshifts the
gas inner radius $r_{g}$ is smaller at the relevant mass ($M_{*}$
is about $10^{14} h^{-1} M_{\odot}$) and is dangerously close to
the effective resolution limit (about $12 h^{-1}$ kpc) of the
simulation. This would tend to reduce the pressure power at small
scale in the simulation and thus explain the discrepancy. Numerical
dissipation could also be at the origin of this small scale power
damping (at a scale comparable to the spatial resolution). Indeed,
this effect tends to induce a spurious increase in entropy, which
results in a larger ``core radius'' in the gas distribution.
Finally, the disagreement could also be caused by the limitations
of the halo model. As Figure~\ref{fig:cluster} illustrates, the
structure of clusters is more complicated than that assumed in our
model. In particular, cluster profiles may not be isothermal. In
fact, our simulations indicate that the gas temperature tends to
rise towards the center of clusters. This would lead to larger
effective core radii and thus tend to suppress power on small
scales.  This effect could thus explain the discrepancy, as it is
more pronounced at low redshifts.

The mismatch of the normalization at small $k$ can be studied using
Figure~\ref{fig:bp_z}, which shows the large-scale pressure bias as
a function of redshift for both methods. The squares show the
predictions for the simulations at the largest available scale,
$k=0.06 h$ Mpc$^{-1}$. The solid line shows the fiducial halo
prediction at the same scale (derived from Eq.~[\ref{eq:bp}]). Both
methods predict that the pressure bias increases with redshift, and
agree relatively well for $z \lesssim 1$. The two methods however
disagree by about 40\% at $z \simeq 3$. In addition, the change of
slope predicted by the halo model at this redshift is not observed
in the simulations.

The large scale pressure bias depends on the mass range chosen for the
halo model. This can be seen by examining the dotted curve in
Figure~\ref{fig:bp_z}. It shows the halo prediction for a wide mass
range, which has reached convergence to the asymptotic limit of an
infinite mass range. This curve and that corresponding to the fiducial
mass range agree at intermediate redshifts ($1 \lesssim z \lesssim
2$), but differ both at lower and higher redshifts: increasing $M_{\rm
max}$ tends to increase $b_{p}$ at low redshift, while decreasing
$M_{\rm min}$ tends to decrease $b_{p}$ at large redshift. While this
dependence must be taken into account, we could not find any
reasonable mass range which eliminated the discrepancy with the
simulations.

The halo prediction for the asymptotic case of $k\rightarrow 0$
(Eq.~[\ref{eq:bp0}]) differs from the $k=0.06 h^{-1}$ Mpc curves shown
in Figure~\ref{fig:bp_z}, only by about 20\%. It therefore captures
the dominating dependence of the large-scale pressure bias, and can
thus be used to understand the origin of the discrepancy. For a given
choice of the mass range, the asymptotic halo prediction is very
robust. Indeed, once the mass function has been fixed to ensure an
agreement with the DM density power spectrum, $b_{p}(0)$ only depends
on the halo bias function $b(\nu)$. After experimenting, we found that
a modification of the Mo \& White \cite{mo96} relation for $b(\nu)$
(Eq.~[\ref{eq:b_nu}]) can indeed reduce the discrepancy. This is
however poorly motivated, and would not be necessarily consistent with
the observed clustering of DM haloes in N-body simulations
\cite{mo96,she99,she00,jin98,rob00}. The Sheth, Mo \& Tormen
\cite{she00} $b(\nu)$ relation was shown to better agree with N-body
results \cite{rob00}; its use in place of the Mo \& White relation
however slightly worsens the discrepancy for the gas pressure. We
therefore do not pursue this possibility here.

Instead, we note that this discrepancy for the large-sale pressure
bias could come from the limitations of the halo model. It indeed
assumes that that all the matter can be modeled as a collection of
collapsed halo. At early times and on large scales, filaments and
sheets indeed dominate the large-scale structure and are awkwardly
described by the halo model. Visual inspection of SZ maps obtained
by the simulation at large redshifts ($z=2$-$3$) confirms this
tendency qualitatively.

Our results for the pressure bias are in good qualitative agreement
with Figure 6 in Refregier et al. \cite{ref00}. They however find a
faster fall off at large $k$ and a larger evolution of the pressure
bias $b_{p}(k)$. These two facts can be explained by the coarser
spatial resolution of their simulations, which, as discussed in
\S\ref{sim_resol}, translates into a larger effective mass resolution
($M_{\rm min} \simeq 5 \times 10^{12} h^{-1} M_{\odot}$).

\subsection{Sunyaev-Zel'dovich Power Spectrum}
\begin{figure}
\centerline{\epsfig{file=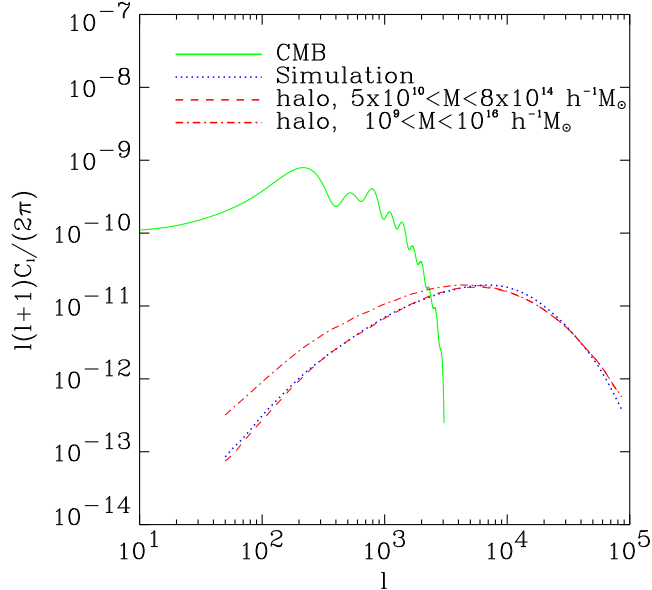,width=90mm}}
\caption{SZ power spectrum for the halo model and for the simulations, in
the RJ regime. The spectrum for the halo model is shown for
different mass ranges: our fiducial mass range (dashed) and the
asymptotic halo model (dot-dashed).  The primordial CMB power
spectrum for the $\Lambda CDM$ model is also shown for comparison.}
\label{fig:cl_m}
\end{figure}

From the estimates of the evolution of the pressure power spectrum
and of the density-weighted temperature, we can compute the SZ
power spectrum, using Equation~(\ref{eq:cl}). The resulting spectra
are shown in Figure~\ref{fig:cl_m} for the simulations, the halo
model with our fiducial mass range, and the asymptotic halo model.
The primordial CMB power spectrum for our $\Lambda$CDM model is
also shown, as derived using {\tt CMBFAST} \cite{zal99}.

The agreement between our fiducial halo model (dashed line) and the
simulations (dotted line) is excellent, showing that some of the
discrepancies in the pressure power spectrum cancel out when
integrated along the line of sight.

The dot dashed line in Figure~\ref{fig:cl_m} shows the halo
prediction for a wide mass range (corresponding to a good
approximation to the asymptotic limit $0<M<\infty$). This model
agrees with our fiducial model (and the simulations) for $\ell
\gtrsim 7000$, but exceeds it by a factor of as much as $\sim 2$
below that. Note that the difference comes primarily from the larger
$M_{\rm max}$ value, while a change in $M_{\min}$ has very little
effect. This shows that the main limitation of the simulation is
the finite box size, which limits the number of massive clusters at
low redshifts.

\begin{figure}
\centerline{\epsfig{file=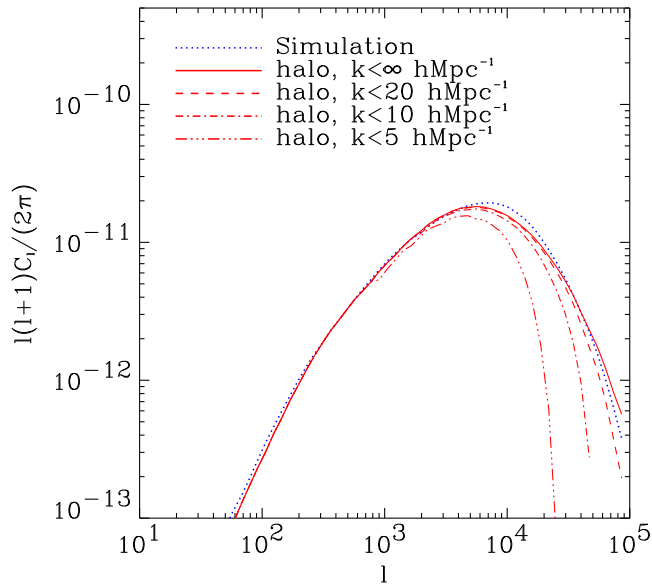,width=90mm}}
\caption{SZ power spectrum for the halo model with different values of
$k_{\max}$. Our fiducial mass limits of $M_{\rm min}=5 \times
10^{10} h^{-1} M_{\odot}$ and $M_{\rm max}=8 \times 10^{14} h^{-1}
M_{\odot}$, and a value of $k_{\min}=0.06 h$ Mpc$^{-1}$ were
adopted. The power spectrum of the simulations is also shown for
comparison.}
\label{fig:cl_k}
\end{figure}

The impact of spatial resolution can be studied by examining
Figure~\ref{fig:cl_k}. For this figure, the SZ power spectrum was
computed from the fiducial halo model, after restricting the
integration of the pressure power spectrum to different upper
bounds $k_{\rm max}$ for the wavenumber $k$.  (In all cases, the
minimum bound was chosen to be $k_{\rm min}=0.06 h$ Mpc$^{-1}$, the
value imposed by the finite size of the simulation box). The figure
shows that the SZ power spectrum is quite sensitive to the spatial
resolution (i.e. to $k_{\rm max}$). A degradation of the spatial
resolution to $k_{\rm max} \simeq 5 h$ Mpc$^{-1}$ leads to a severe
drop of power for $\ell \gtrsim 2\times 10^{3}$ (see also
\cite{ref00}). Interestingly, the SZ spectrum does not vary much if
$k_{\rm max}$ is taken be 20 $h$ Mpc$^{-1}$ or above. For our
simulation, $k_{\rm max} \simeq 60 h$ Mpc$^{-1}$ at low redshifts,
which is sufficient to reach convergence.  The main limitation of
our prediction from the simulation is thus that arising from the
finite box size. On the other hand, thanks to the AMR scheme,
spatial and mass resolution are not a limiting factor.

Our prediction for the SZ power spectrum agrees approximately with
the results of da Silva et al. \cite{das00}, up to a small overall
rescaling due to our larger value of $\sigma_{8}$. Our results also
agree with the prediction of Refregier et al. \cite{ref00}, within
their $\ell$-range of confidence. As they noted, their predictions
were limited by the spatial resolution of their simulations, which
corresponds to $k_{\rm max} \approx 5 h$ Mpc$^{-1}$. Our curve in
Figure~\ref{fig:cl_k} corresponds to this value of $k_{\rm max}$
and is in good agreement with their result.

\section{Conclusions}
\label{conclusion}
We have studied the 2-point statistics of the gas pressure and of the
resulting SZ fluctuations using two methods: a RAMSES numerical
simulation, using Adaptive Mesh Refinement, and an analytical halo
model. Overall, the agreement is good, once the mass resolution and
finite box size of the simulations are accounted for. The agreement is
almost surprising, given the simple nature of the halo model.  The
halo model\cite{coo00} is self-consistent and relies on the assumption
of virial and hydrostatic equilibrium of the gas. After it has been
tuned to produce the correct DM density power spectrum, the model does
not require any new free parameters and thus has a strong predictive
power.

The predictions for the SZ power spectrum from our numerical
simulations are mainly and solely limited by the finite box size.
The AMR techniques indeed gives a reliable dynamic range of about 3
orders of magnitude for the power spectra, corresponding to a
formal resolution of 8192 in linear scale, relieving us from the
limiting effect of finite spatial resolution. The halo model allows
us to extrapolate to an infinite box size, thus providing us with
an accurate SZ power spectrum in the full multipole range $10^{2}
\lesssim \ell \lesssim 10^{5}$. This is an important improvement
upon earlier calculations \cite{ref00,sel99} which were severely
affected for $\ell \gtrsim 2\times10^{3}$ by their coarser spatial
resolution.

In principle, the effect of finite box size could be alleviated
using larger simulations, as discussed in \S\ref{sim_resol}.
However, to achieve convergence, one would need to reach a limiting
mass of $M_{\rm max}= 2 \times 10^{15} h^{-1} M_{\odot}h^{-1}$
which would correspond to a box size of 400 $h^{-1}$ Mpc in a
$\Lambda$CDM model. Given our current computer resources, this is
prohibitive for the near future. One thus has to rely, as we have
done, on analytical models to extrapolate beyond the sampling
variance of the simulations. Analytical models, like the one we
have presented here, also have the advantage of providing a fast
way to explore a wide range of cosmological parameters. They also
help to provide a physical understanding of the physics and scales
which contribute to the SZ statistics. Using our halo model, we
have shown for example that two characteristics length scales
contributes to the SZ power spectrum: the virial radius $r_v$ and
the gas inner radius $r_g$. It would be interesting to understand
in more details how the SZ power spectrum provides a measure of the
different regions of the NFW profile on a range of mass scales.
Another interesting avenue is the study of processes such as
feedback and cooling on the SZ statistics. This will be explored in
future work.

\acknowledgments We thank Francois Bouchet for useful discussions. AR
is grateful for the hospitality of Marguerite Pierre and the
Service d'Astrophysique at CEA/Saclay, where part of this work was
conducted. AR was supported by a TMR postdoctoral fellowship from
the EEC Lensing Network, and by a Wolfson College Research
Fellowship. Simulations were performed on the VPP 5000 Fujistu
computer at CEA Grenoble. RT thanks Philippe Kloos for helping him
on optimizing the RAMSES code.


\end{document}